\documentstyle[11pt,aaspp4]{article}

\def\etal{{\it et al.\/ }}
\def\kms{$\,$km$\,$s$^{-1}$}

\lefthead{Shen et al.}
\righthead{Southern Hemisphere VLBI Survey - I}

\begin{document}
\title{A 5-GHz Southern Hemisphere VLBI Survey of \\
               Compact Radio Sources - I}
\vspace{16pt}
\author{Z.-Q. Shen}
\affil{Harvard-Smithsonian CfA, 60 Garden Street, Cambridge, Massachusetts 02138; and\\
       Shanghai Astronomical Observatory, 80 Nandan Road, Shanghai 200030, P.R.China}

\author{T.-S. Wan}
\affil{Shanghai Astronomical Observatory, 80 Nandan Road, Shanghai 200030, P.R.China}
 
\author{J. M. Moran}
\affil{Harvard-Smithsonian CfA, 60 Garden Street, Cambridge, Massachusetts 02138}

\author{D.~L.~Jauncey, J.~E.~Reynolds, A.~K.~Tzioumis, R.~G.~Gough, R.~H.~Ferris, and M.~W.~Sinclair}
\affil{Australia Telescope National Facility, CSIRO, Epping, NSW~2121, Australia}
 
\author{D.-R. Jiang, X.-Y. Hong, and S.-G. Liang}
\affil{Shanghai Astronomical Observatory, 80 Nandan Road, Shanghai 200030, P.R.China}

\author{M. E. Costa}
\affil{Physics Department, University of Western Australia, Nedlands, WA 6009, Australia}

\author{S. J. Tingay}
\affil{Mount Stromlo and Siding Spring Observatories, ACT 2611, Australia}

\author{P. M. McCulloch, J. E. J. Lovell, and E. A. King}
\affil{Department of Physics, University of Tasmania, Hobart, Tasmania 7001, Australia}

\author{G. D. Nicolson}
\affil{Hartebeesthoek Radio Astronomy Observatory, Krugersdorp 1740, South Africa}

\author{D. W. Murphy, D. L. Meier, and T. D. van Ommen}
\affil{Jet Propulsion Laboratory, California Institute of Technology, Pasadena, California 91109}

\author{P. G. Edwards\altaffilmark{1}}
\affil{Department of Physics, University of Adelaide, Adelaide, SA 5005, Australia}

\author{G. L. White}
\affil{Physics Department, University of Western Sydney, Nepean, NSW 2747, Australia}

\altaffiltext{1}{Present address: The Institute of Space and Astronautical Science, Sagamihara,
                           Kanagawa 229, Japan}
\clearpage

\begin{abstract}
We report the results of a 5-GHz southern hemisphere VLBI survey
of compact extragalactic radio sources. 
These observations were undertaken with the SHEVE array plus Shanghai
station in 1992 November. 
A sample of twenty-two sources was observed and images of twenty of them were obtained.
Of the twenty sources imaged, fifteen showed core-jet structure, one had a
two-sided jet, and four had only single compact cores. 
Eleven of the sixteen core-jet (including one two-sided jet) sources showed
some evidence of bent jets. 
No compact doubles were found. 
A comparison with previous images and the temporal variability of the
radio flux density showed evidence for superluminal motion 
in four of the sources.
Five sources were high-energy ($>$~100~MeV) $\gamma$-ray sources.
Statistical analysis showed the dominance of highly polarized quasars among
the detected $\gamma$-ray sources, which emphasizes the importance of
beaming effect in the $\gamma$-ray emission.
\end{abstract}

\section{Introduction}
While considerable VLBI imaging and monitoring of extragalactic radio sources have
been carried out in the northern hemisphere, the only VLBI surveys
undertaken in the south have been the systematic
one-baseline intercontinental surveys at 2.3~GHz (Preston \etal 1985)
and 8.4~GHz (Morabito \etal 1986) and the 2.3~GHz imaging and
model-fitting SHEVE (Southern Hemisphere VLBI Experiment) 
survey of twenty-nine sources in 1982 (Preston \etal 1989;
Meier \etal 1989; Tzioumis \etal 1989; Jauncey \etal 1989).

In the north, Pearson and Readhead (1981, 1988) conducted a pioneering
systematic VLBI survey that produced images of thirty-seven strong extragalactic
sources from a complete sample of sixty-five sources. 
Most recently, the first
Caltech-Jodrell Bank VLBI Survey (CJ1) (Polatidis \etal 1995;
Thakkar \etal 1995; Xu \etal 1995) 
and the second Caltech-Jodrell Bank VLBI Survey (CJ2) (Taylor \etal 1994;
Henstock \etal 1995) imaged 135 and 193 sources respectively.
These programs have produced a systematic classification scheme for compact
extragalactic radio sources.

The launch of the Space VLBI missions VSOP (Hirosawa 1991) in 1997 and
the intended launch of RadioAstron
(Kardashev \& Slysh 1988) will lead to a significant increase in
resolution over that possible with ground-based radio telescopes alone. 
In order to make full use of the resources of Space VLBI, preliminary
ground-based surveys on intercontinental baselines are essential to 
identify suitable targets.

With these imperatives to improve the understanding of compact radio sources in
the south, we have been conducting a 5-GHz southern hemisphere VLBI radio
source survey, with the major aims of: (1) filling the gap in southern VLBI
observations; (2) presenting a pre-launch survey for those potential targets
for space VLBI; (3) providing first-epoch observations in a search for
southern superluminal sources; and (4) identifying compact sources as
potential southern calibration and astrometric reference sources. 
We have undertaken two observing sessions.  The first, described here, 
was completed in November 1992 with the observation of twenty-two sources.  The second 
was undertaken in May 1993, when a further twenty-three sources were observed. 
 
This paper describes the astrophysical results of the first period of 
observations. We introduce the survey sample  ($\S$2); 
describe the observations and data reduction procedures ($\S$3); 
present the results with emphasis on the twenty sources imaged ($\S$4); 
analyze VLBI measurements in the light of the detection of active galatical
nuclei by EGRET (Energetic Gamma-Ray Experiment Telescope) on the
CGRO (Compton Gamma-Ray Observatory) ($\S$5);
and summarize our conclusions ($\S$6).

\section{The Sample}
We selected those southern radio sources with the 
strongest correlated flux densities at frequencies near 5~GHz 
to help ensure successful imaging surveys. 
These extragalactic sources were chosen from the one-baseline intercontinental surveys 
at frequencies of 2.3 and 8.4~GHz (Morabito \etal 1986 and references therein) 
using the following selection criteria:
 
\begin{enumerate}
\item{ declination: $-45^\circ$$<$$\delta$$<$$+10^\circ$, }
\item{ correlated flux densities at 2.3 and 8.4~GHz: 
S$^{\small c}_{\small 2.3GHz}$$>$0.6~Jy and 
S$^{\small c}_{\small 8.4GHz}$$>$0.6~Jy, }
\item{ total flux density at 5~GHz: S$^{\small t}_{\small 5GHz}$$>$1.0~Jy. } 
\end{enumerate}
 
\noindent
Thirty-six sources meet these criteria, including the
extensively observed and well-studied sources 3C 273 and 3C 279.
These sources are, in order of their right ascensions: 
0048$-$097, 0104$-$408, 0106+013,
0332$-$403, 0420$-$014, 0426$-$380, 0438$-$436, 0458$-$020, 0521$-$365,
0727$-$115, 0736+017, 1034$-$293, 1055+018, 1104$-$445, 1144$-$379,
1226+023, 1253$-$055, 1334$-$127, 1354$-$152, 1504$-$166, 1510$-$089,
1519$-$273, 1548+056, 1730$-$130, 1741$-$038, 1749+096, 1921$-$293,
2121+053, 2131$-$021, 2134+004, 2145+067, 2216$-$038, 2223$-$052,
2243$-$123, 2318+049, and 2345$-$167.
This sample is a subset of the primary list proposed by the 
Science Objectives Committee for observation by RadioAstron,
which has a major selection criterion of correlated flux densities 
in excess of 0.5~Jy at 2.3 or 8.4~GHz.
We chose twenty of the thirty-six sources for the first experiment,
due to the limited observing time available.
In addition, we also added two other interesting sources, 
0235+164 (well-studied, and a member of the RadioAstron list,
but outside the declination range of the criteria)
and 1814$-$637 (strong, but also outside the declination range).
All twenty-two sources could be treated as Parkes sources.
They have been optically identified, and redshifts are
available for twenty of them.
Their names, positions, redshifts, optical types, flux densities 
and related information are listed in Table 1.
 
Of the remaining sixteen sources conforming to the above selection criteria,
seven (0332$-$403, 0426$-$380, 0438$-$436, 0521$-$365, 1034$-$293, 1226+023 and
2243$-$123) were observed in May 1993.
The other nine sources (0727$-$115, 0736+017, 1055+018, 1253$-$055, 1354$-$152, 
1749+096, 2121+053, 2131$-$021 and 2318+049) will be observed
in future experiments.
The results of these experiments will be reported later.

\section{Observations and Data Reduction}
We performed the survey with the SHEVE 
network, augmented by the Shanghai station, on 1992 November 20--22.
The participating stations were located in Hartebeesthoek (South Africa), 
Hobart (Australia), Mopra (Australia), Narrabri (Australia), Perth (Australia)
and Seshan (Shanghai, China).
The station parameters are listed in Table 2.
Most of the twenty-two radio sources in Table~1 were observed in 
a snapshot mode,
i.e., two to five thirty-minute scans were obtained.
Figure 1 shows a typical (u,v) plot for 1730$-$130 with three scans in the observations.
At the beginning and/or the end of each scan, system temperature measurements 
for the amplitude calibration were made and the antenna pointing was checked.
All data were recorded in VLBI Mark~II format with 2-MHz bandwidth and
left-circular polarization (IEEE convention) at each station during the observation,
and subsequently cross-correlated on the JPL/Caltech Mark~II processor 
in Pasadena, California, in September 1994.
 
The post-correlation reduction was carried out at
the Harvard-Smithsonian Center for Astrophysics,
using the NRAO
AIPS and Caltech VLBI analysis packages.
We first applied the AIPS global fringe-fitting algorithm.
The raw data was edited based on 
information contained in the logsheets from each station.
Then {\it a~priori\/} amplitude calibrations were applied using the gain curves and 
system temperatures measured at each station.
In the fringe-fitting procedure we used a point-source model and
a solution interval of 2.5 minutes.
Hobart served as the reference telescope whenever possible.
Fringes were found for each source 
on almost all the baselines. 

The AIPS-format data were then converted to the Caltech format for mapping
with the program DIFMAP (Shepherd, Pearson, \& Taylor 1994)
in the Caltech VLBI Package.
The 2-second visibility data were coherently time-averaged to 60 seconds.
The uncertainties in the averaged visibilities were computed from the scatter
of data points within the averaging interval.
The data were inspected for obviously bad points, most of which were near
the start time of a scan when telescopes were still slewing to the sources, 
and these were removed.
The sources were mapped with several iterations of the self-calibration
and cleaning procedures.
Natural weighting of the data was used for all imaging.
A 1$-$Jy point-source model was employed at the start of each mapping process.
On subsequent iterations, the image produced by previous clean cycles was used
as an input model for phase corrections.
For the amplitude self-calibration, a constant 
gain factor for each antenna was implemented in the later stage of 
the mapping.
The derived factors for each antenna were usually in good agreement 
with those obtained by comparing the correlated flux densities on different 
baselines at the crossing points in the (u,v) plane, where available.
Twenty maps were successfully produced for this survey 
(see Figure 2 and Table 3).

We analyzed the images quantitatively using
the MODELFIT program in the Caltech VLBI package.
Up to three Gaussian components were used to fit both the closure phases and amplitudes 
in the calibrated data.
The results are listed in Table~4.
In some cases, the model does not fit the data as precisely 
as expected based on a $\chi^2$ criterion, indicating 
that the source distribution is probably more complex than 
our model.
Generally speaking, the models are not unique, and it is possible to find 
other parameters which yield an acceptable fit to the data.
We have generated a family of acceptable models for each source, and found 
that these do not differ significantly from our adopted models.
We have also done Monte Carlo simulation and demonstrated that all the model
parameters are robust, with the exception, in some cases, of the component position angle
of the major axis.
The uncertainties in position angle are large when the signal-to-noise ratio
is low (i.e., the component is weak) and/or the axial ratio is close to unity.
In all cases we feel that the models given are reasonable representations 
of the sources.

\section{Image and Discussion of Individual Sources}

We discuss the results of each source in our survey in order of
right ascension.
The maps of the sources are shown in Figure~2.
On each map, the size of the restoring beams is shown as a cross-hatched 
ellipse in the lower left corner.
The lowest contour level in each map is three times the rms noise.
The rms noise in the images varied from 5~mJy to 10~mJy per beam for
most sources.
The values of various parameters for each source (e.g., peak flux density,
rms noise,
restoring beam and contour levels) are listed in Table 3.
The results of the model-fitting are given in Table 4.
 
It is useful to make a quantitative determination about whether a source 
component
is resolved (i.e., shows signs of being extended in angle)
or unresolved by the interferometer.
We established a criterion based on the following analysis.
The visibility of a circular Gaussian component is
\begin{equation}
V~=~V_0~e^{-3.56(\frac{\theta_{\small S}}{\theta_{\small B}})^2},
\end{equation}
where V$_0$ is a constant equal to the flux density value for
an unresolved source,
while $\theta_{\small B}$ is the resolution (or synthesized beam width) and
$\theta_{\small S}$ is the source size in full width to half maximum intensity (FWHM).
It is clear that the source is resolved if $\theta_{\small S}$~$>$~$\theta_{\small B}$.
Therefore, we only consider the case of $\theta_{\small S}$~$<$~$\theta_{\small B}$.
From equation (1), we can derive the fractional deviation of visibility as
\begin{equation}
\frac{\Delta{V}}{V_0}~=~3.56~
(\frac{\theta_{\small S}}{\theta_{\small B}})^2,
\end{equation}
where $\Delta{V}=V_0-V$ is the 
decrease in visibility due to resolution.
Instrumental contributions to 
the fractional deviation, $\frac{\Delta{V}}{V_0}$,
might be caused by two terms:
a statistical term which is the reciprocal of 
the signal-to-noise ratio (SNR), and a systematic one, 
$\frac{\Delta{F}}{F}$, which is the sum of the uncertainties in the 
flux density measurements and calibration errors,  and is given by
\begin{equation}
\frac{\Delta{V}}{V_0} ~=~\left\{
{(\frac{1}{\normalsize SNR})}^2~+~
{(\frac{\Delta{F}}{F})}^2
\right\}^
{1/2}.
\end{equation}
Equating the right-hand sides of Equations (2) and (3), we get
\begin{equation}
\theta_{\small LIM}~=
\left\{\theta_{\small LIM}(\normalsize Statistical)^4~+~
\{\theta_{\small LIM}(\normalsize Systematic)^4\right\}^{1/4},
\end{equation}
where $\theta_{\small LIM}$ is the limit 
on source size that can be set, 
and $\theta_{\small LIM}(\normalsize Statistical)$
and $\theta_{\small LIM}(\normalsize Systematic)$ are given by
\begin{equation}
\theta_{\small LIM}(\normalsize Statistical)~=~0.53~
\frac{\theta_{\small B}}{\sqrt{\normalsize SNR}}~~,
\end{equation}
and 
\begin{equation}
\theta_{\small LIM}(\normalsize Systematic)~=~0.53~
\frac{\theta_{\small B}}{\sqrt{\left| F/\Delta{F} \right|}}~~.
\end{equation}
For our data, $\frac{\Delta{V}}{V}$ is usually dominated by 
the systematic errors.
If 
$\left| \frac{\Delta{F}}{F} \right|~\sim~0.15$ (a reasonable estimate for our
data) and SNR~$\sim$~10, then 
\begin{equation}
\theta_{\small LIM}(\normalsize Statistical)~=~0.17~{\theta_{\small B}}
\end{equation}
and
\begin{equation}
\theta_{\small LIM}(\normalsize Systematic)~=~0.21~{\theta_{\small B}}.
\end{equation}
Since SNR$>$10 in all cases in Table 4, $\theta_{\small LIM}$ is
always dominated by the systematic term (see equations (4), (7) and (8)), 
and we adopt
\begin{equation}
\theta_{\small LIM}~=~0.2~{\theta_{\small B}},
\end{equation}
where we take $\theta_{\small B}$ to be the geometric mean of the major 
and minor beam sizes.
The results based on this criterion are listed in column 10 of Table 4.
Components with fitted sizes smaller than $\theta_{\small LIM}$ in Equation (9)
are listed as unresolved.

The peak brightness temperature of each component 
in the rest frame of the source is
listed in column 9 of Table 4, 
and was calculated from the relation
\begin{equation}
{\normalsize T}_{\small b}~=~1.22~10^{12}~{\normalsize S}~\nu^{-2}~
(\theta_{\small max}~\theta_{\small min})^{-1}~(1+{\normalsize z})~
{\normalsize K}~~,
\end{equation}
where S is the flux density in Jy, $\nu$ the frequency in GHz (5~GHz in our
experiment), z the redshift, and $\theta_{\small max}$ and 
$\theta_{\small min}$ are the major and minor axes in mas, respectively.
For unresolved components, lower limits are provided.

Throughout the paper, we define the spectral index, $\alpha$, 
by the convention S$_{\small \nu}$$\propto$${\nu}^{\alpha}$, and we
assume H$_0=100$ \kms~Mpc$^{-1}$ and q$_0=0.5$.
Our search of the literature was greatly assisted by NED
\footnote {The NASA/IPAC Extragalactic Database~(NED) is operated by 
the Jet Propulsion Laboratory, California Institute of Technology, 
under the contract with the National Aeronautics and Space Administration.}.
\\
 
\noindent
{\bf 0048$-$097 (OB$-$080, Fig.~2A )}
 
This is a bright, strongly variable BL Lac object (Andrew \& Smith 1983).
The absence of the extended nebulosity of the BL Lac
host galaxy suggests that z$>$~0.2
(Stickel, Fried, \& K\"{u}hr 1993).
It is highly variable at mm wavelengths 
(Steppe \etal 1988; Steppe \etal  1992), 
which is consistent with its flat spectrum
between 8.4 and
90~GHz (Tornikoski \etal 1993).

No previous VLBI images have been obtained at any frequency.
From earlier VLBI observations at 5~GHz, Weiler and Johnston (1980) estimated 
the visibility to be 0.89 with total flux density of 0.80~Jy,
and Gaussian model diameter of 0.4~mas.

Our observations revealed a core-jet structure 
extending 2.0~mas from the central core
at a position angle (P.A.) of $-159^{\circ}$. 
This orientation is consistent with the large core-jet structure 
on the scale of 7~arcseconds observed with the VLA
(Perley 1982; Wardle, Moore, \& Angel 1984).
The parameters of the compact core (component 1) are in reasonable 
agreement with the results from Weiler and Johnston (1980) in both visibility 
and source size,
although the flux density has increased about 1.5 times. 
There is no indication in the radio data 
of the elongation to the east,
which was once seen on an optical CCD image (Falomo, Melnick, \& Tanzi 1990),
but not detected in the subsequent spectroscopic observations
(Stickel, Fried, \& K\"{u}hr 1993).\\

\noindent 
{\bf 0104$-$408 ( Fig.~2B ) }
 
This 19th-magnitude QSO was identified as a possible BL Lac object with
a redshift z=0.584 (White \etal 1988).
It is one of the benchmark objects in the
establishment of a high-precision radio/optical celestial reference frame
(Costa \& Loyola 1992).
There are no previous radio images of the source.
 
The source is well-fitted by a single Gaussian component with a flux density
of 2.6~Jy and a size of 0.6~$\times$~0.3~mas at P.A.~=~$19^{\circ}$.
It might have a slight extension toward the east.
Because of its southern declination, 
the Shanghai antenna did not contribute to the mapping, which caused
a relatively poor synthesized beam (2.8~$\times$~0.7~mas at P.A.~=~$4^{\circ}$) 
in the north-south direction.
The total flux density was 2.6~Jy, about three times stronger than 
at previous epochs (e.g. Quiniento, Cersosimo, \& Colomb 1988).
Future observations at higher resolution (such as space VLBI) would be useful
to probe the central core.
\\

\noindent
{\bf 0106+013 (4C~01.02; OC~012; Fig.~2C )}

\nopagebreak
This source is a highly polarized quasar, and with z=2.107 (Burbidge 1966),
has one of the highest redshifts in our sample.
Previous two-epoch VLBI observations were made by
Wehrle, Cohen, \& Unwin (1990b).
Their 5-GHz maps showed an east-west structure and a relative component 
proper motion of $\mu=0.2$~mas~yr$^{-1}$, corresponding to an apparent 
superluminal motion with $\beta_{\small app}=8.2$. 
We fit new data to a model consisting of three Gaussian components.
Their parameters are listed in Table 4.

The VLA image at 5~GHz exhibits a weak jet 5~arcseconds long
extending to the south (Kollgaard, Wardle, \& Roberts
1990).
Thus, the jet bends through an angle of $90^\circ$ moving from the VLBI-core 
to the outer region.
Neff and Hutchings (1990) estimated the bending angle to be $85^\circ$ 
without detailed justification.
Our image, with  its good 
resolution in the north-south direction,
clearly shows the curvature in the jet
on the mas scale.
We note that a high-redshift source is more likely to show a bent jet 
than a source at low redshift (Barthel \& Miley 1988).
Higher dynamic-range imaging will help us better define this
morphological transition.

Wehrle, Cohen, \& Unwin (1990b) could not identify the core 
because of the lack of spectral 
index information.
Assuming that the relative proper motion is constant (0.2 mas~yr$^{-1}$)
along the direction of P.~A.~$\sim$~$-110^\circ$, 
we estimate the separation to be 2.03~mas in November 1992.
This is consistent with the distance on our map 
between components 1 and 3,
which are separated by 2.06~mas along a P.A. of $-116^\circ$. 
However, the flux density of component 3 has decreased dramatically 
to 0.20~Jy compared with 2.8~Jy in 1986.
As for component 1, it was weak in 1986 (0.8~Jy), 
peaked around 1988 (1.8~Jy), and decreased to 1.5~Jy by 1992.
This modest variability suggests that component 1 
is most probably the core. 
A dramatic variation has been observed 
in the flux density within the region of 1~mas in 1984$-$1988
(Takahashi \& Kurihara 1993).
This may be related to 
a new bright component emerging from the core.
The peak variation around 1987.8 is probably connected with 
the emergence of component 2.
This implies a proper motion of 0.18~mas~yr$^{-1}$, or a superluminal
motion of 7.4c, for component 2.
\\

\noindent 
{\bf 0235+164 (OD~160; Fig.~2D )}
 
This widely studied object
is one of the most violently variable sources known
(Pollock \etal 1979).
It is identified with a BL Lac object (Spinrad \& Smith 1975),
and has three distinct redshifts, one in emission at z=0.940 (Cohen, Smith,
\& Burbidge 1986), one in absorption at z=0.851 (Burbidge \etal 1976), and one
in both emission and absorption at z=0.524 (Rieke \etal 1976).
 
In spite of its complex optical features, the radio morphology of
0235+164 is surprisingly simple on both arcsecond and mas scales.
The VLA maps at 1.4~GHz and 5~GHz showed no extended emission 
above 0.1\% of the peak flux density of the core 
(Ulvestad, Johnston, \& Weiler 1983; Antonucci \& Ulvestad 1985).
In the VLBI maps, most of the flux density comes from 
the central compact core (Jones \etal 1984).
No mas polarization structure was found \hbox{at 5~GHz (Gabuzda \etal 1992).}

It has been shown that each strong optical 
and radio outburst is accompanied by the formation of a new VLBI component
(e.g., B{\aa}{\aa}th 1984).
These components extend to the northeast, but the position
angles vary between $10^\circ$ and $45^\circ$  among epochs.
Our map, which agrees with such a description, 
shows a very compact core and a weak shoulder to the
northeast (see Table 4).
The total flux density measured independently at the time of our experiment at 
Hartebeesthoek was 4.3~Jy, higher than the typical value ($\sim$3.0~Jy).
The brightness temperature of the central core exceeds the 
Compton limit by a factor of 5 for a redshift of 0.94. 
\\

\noindent 
{\bf 0420$-$014 (OA~129; Fig.~2E )} 

The 5-GHz image of this flat-spectrum radio quasar exhibits a resolved core (component 1) of 
2.0~Jy and a strong secondary component (2) of 1.7~Jy 
along the direction of $-146^\circ$ at a distance of 0.96~mas from 
the core. 
The total flux density of 0420$-$014 obtained from Hartebeesthoek is about 
4.0~Jy, which is less than that observed by Wehrle \etal (1992) in 1986.44. 
From the radio flux-density curve available (Wehrle \etal 1992), 
the total flux density of 
0420$-$014 reached a peak around 1986.50 at 14.5~GHz, and 1987 at 
5.0~GHz. 
Such a peak may be related to the emergence 
of a new component, which could explain the tight core-halo structure mapped 
by Wehrle {\it et al.}, if the jet lay very close to the line of sight. 
Assuming that component 2 was ejected at the 
time of the measurements made by Wehrle \etal (i.e. 1986.44), 
we derive a proper motion of 0.15~mas~yr$^{-1}$, 
corresponding to 
an apparent transversal speed of 3.9c at redshift z=0.915. 
This is not surprising if we consider that 
0420$-$014 is one of the $\gamma$-ray sources 
detected by EGRET on the CGRO (von Montigny \etal 1995; Radecke \etal 1995). 
Moreover, in the study of geodetic VLBI experiments, Wagner \etal (1995)
reported a p²oper motion of 0.15~mas~yr$^{-1}$ along a line of P.A.
$-142^\circ$ at the epoch of 1992.15. The jet motion might relate to the
southern extended emission revealed by 1.4~GHz VLA observations (Antonucci \&
Ulvestad 1985). Further VLBI observations are still needed to confirm these
results and study the motion of the jet component.
\\

\noindent 
{\bf 0458$-$020 (4C$-$02.19; Fig.~2F )}
 
This source has the highest redshift (z=2.286) both in our sample
and in the sample of  active galactic nuclei detected
by EGRET in high-energy $\gamma$-rays (E$>$100~MeV) 
(von Montigny \etal 1995).
Its radio structure extends over a wide range of scales.
The VLA observations at 1.4~GHz (Briggs \etal 1989) show two distinct
components (1.8~arcseconds in separation with a P.A. of $-127^\circ$).
The VLBI map at 608~MHz (Briggs \etal 1989)
shows a jet that heads northwest initially and 
then veers to the southwest.
The position angle of the innermost structure is $-58^\circ$.
 
Our 5-GHz map confirms the core-jet structure of the source, which was first
observed by Wehrle \etal (1992). 
Our map can be represented by three components:
a strong compact core (component 1) of 2.6~Jy,
and two other resolved weak components (2 and 3).
Table 4 describes these components in detail.
The detailed description of these components is listed in Table 4.
There probably exists a bending angle of $\sim80^\circ$ from component 2
to component 3, which then connects to the curvature observed by the VLA
(Briggs \etal 1989).
This will be defined better by higher dynamic range observations.
 
Comparing the Wehrle \etal map (1989.8) with our map, we find that component 3 
matches the jet component identified by Wehrle \etal
Component 2 has no obvious counterpart in the Wehrle \etal map,
perhaps because of their poorer resolution in the south-north direction.
No sign of motion in the jet could be measured.
Vermeulen \& Cohen (1994) also did not detect proper motion.
\\

\noindent 
{\bf 1104$-$445 (Fig.~2G )}
 
Partly because of its southern declination, 1104$-$445 has been less well
studied with VLBI, although it is very strong at 5~GHz 
(3.1~Jy at the time of our experiment).
Due to the lack of data for longer baselines to Hartebeesthoek,
Preston \etal (1989) did not resolve its fine structure at 2.3~GHz
with SHEVE in 1982.
 
The source has a core-jet structure on our map.
Its structural parameters  are listed in Table 4.
There is evidence of jet curvature at a distance of $\sim$~1.8~mas 
moving outward from the core (corresponding to a projected length of $\sim8.0~$pc)
from the northeast to the north.
The lack of baselines to the Shanghai antenna caused a relatively poor  
resolution in the north-south direction  (2.6~$\times$~1.0~mas at P.A.~=~$13^\circ$ of beam).
Further VLBI imaging observations are needed to monitor the changes in the
jets.
\\

\noindent 
{\bf 1144$-$379 (Fig.~2H )}
 
This is a BL Lac object (Nicolson \etal 1979) 
with the highest emission-line redshift (z=1.048) in the 1$-$Jy
sample (Stickel, Fried, \& K\"{u}hr 1989; Stickel \etal 1991).
Considering its high optical variability and polarization,
Impey and Tapia (1988) classified it as a blazar.
There is no previous VLBI map.

This highly variable radio source was in an active phase at the time 
of the observations with a flux density at 5.0 GHz of about 4.6~Jy.
Our observations, without baselines to the Shanghai antenna,
showed it as unresolved.
The visibility data can be fitted by a 0.5~mas~$\times$~0.4~mas
Gaussian component with 4.6~Jy of flux density, which corresponds to a
brightness temperature of 2.3~$\times$~$10^{12}$ K.
Higher-frequency VLBI observations are important for the
investigation of the structural variation of this source.
\\

\noindent 
{\bf 1334$-$127 (PKS~1335$-$127; OP$-$158.3; Fig.~2I )}
 
Based on the optical polarimetry, 
1334$-$127 was classified as a highly polarized quasar, or a blazar 
(Impey \& Tapia 1988, 1990). 
VLA observations at 1.4~GHz showed a 
curved jet extending to 6.5~arcseconds east of the core (Perley 1982). 
Wehrle \etal (1992) found, from VLBI observations in 1986.9, 
that the source was barely resolved.
 
This source can be represented by a core-jet structure from our 
observations (see Table 4).
The weak jet component is located at 1.7~mas to the core 
along a direction of $-160^\circ$, 
and must bend through about $110^\circ$ between mas
and arcsecond scales.
1334$-$127 was at its peak 
flux density in 1986.9, but was not very active during our observations. 
It is clear that the source was undergoing an outburst during the 
observations of Wehrle \etal (1992) based on the monitoring observations 
at 14.5~GHz and 8.4~GHz.
Such an outburst would be first detected at higher frequencies and then followed
at lower frequencies, which could explain the failure of Wehrle \etal
to detect the jet component at 5~GHz.
Assuming the jet component was ejected in 1986.9 with the outburst, we 
estimate the average transverse angular velocity to be 
0.28~mas~yr$^{-1}$, 
which corresponds to $\beta_{\small app}=5.2$
at a redshift of 0.541 (Wilkes 1986).
Hence, 1334$-$127 is a superluminal candidate.
\\

\noindent 
{\bf 1504$-$166 (PKS~1504$-$167; OR$-$102; Fig.~2J )}
 
This source shows low-frequency variability (McAdam 1982;
Bondi \etal 1994).
It is also classified as a highly polarized quasar or blazar 
(Impey \& Tapia 1988, 1990).
The 1.7~GHz VLBI two-epoch maps (Romney \etal 1984; 
Padrielli \etal 1986) revealed a very compact structure in 1504$-$166, 
with a jet component at about 3.5~mas from the compact core towards
the direction of $140^\circ$.
 
Our map consists of a circular compact core (component 1)
with a flux density of 1.3~Jy and a diameter of 0.5~mas, 
a 0.6~Jy jet (component 2) at the position angle of $-156^\circ$ 
and separation of 1.12~mas to the core, and 
a possible 0.3~Jy component (3) at 0.80~mas to the core with 
a P.A. of 161$^\circ$.
The third component is close in position angle, 
but has a different scale-size compared to 
the jet component at 1.7~GHz (Padrielli \etal 1986).
All these components are resolved.
The difference of about $45^\circ$ between the position angles of the two
components (2 and 3) within 1$-$2~mas from the core would be interesting
for future observations.
\\

\noindent 
{\bf 1510$-$089 (OR$-$017; Fig.~2K )}
 
1510$-$089 has an optical redshift of 0.361 (Burbidge \& Kinman 1966),
one of the lowest in our sample.
It is a core-dominated, highly polarized quasar (Moore \& Stockman 1981)
with low-frequency variability (Padrielli \etal 1987;
Ghosh, Gopal-Krishna, \& Rao 1994).
1510$-$089 is one of the $\gamma$-ray sources detected by EGRET on the CGRO 
(Thompson \etal 1993), and is so compact and strong that 
fringes have been detected on baselines of 15,000~km
by the TDRSS 15-GHz space VLBI (Linfield \etal 1990).
 
Our observations of 1510$-$089, with an improved north-south resolution,
confirmed the asymmetric two-sided structure at 5~GHz (de Waard 1986).
Our analysis shows that two of three components (1 and 2) are unresolved (see Table 4).
In addition, there is probably weak emission to the south, which 
may be part of the jet feature extending 4~mas in length in
P.A. $\sim173^\circ$ observed by Romney \etal (1984),
and which may also be connected to the arcsecond-scale jet
component extending 8 arcseconds at a P.A. of $160^\circ$ (Perley 1982).
The two-sided features, components 2 and 3 on our map,
are asymmetric in flux density and in their positions 
relative to the central component.

Comparing our results with those obtained in 1984.25 at 5~GHz by de Waard (1986),
we conclude that the fine structure of the source at 5~GHz has not changed 
significantly
over this eight-year period.
The northwest component (3) 
could be related to the increase in angular 
size of the compact component at 1.7~GHz in a P.A. of $-10^\circ$ 
reported by Padrielli \etal (1986).
The VLA images at 15 and 23~GHz (O'Dea, Barvainis, \& Challis 1988) 
showed a slightly resolved component 300~mas from the core at 
a position angle of  
$-28^\circ$.
It is possible that the northwest component in our map extends from
the core initially at a P.A. of $-42^\circ$ and then curves north by 
$\sim15^\circ$ to point at the 300-mas secondary component.
The southeast feature (component 2) could connect to the southern arcsecond 
structure revealed by many VLA observations
(e.g. 1.4 and 5~GHz by Price \etal 1993;
5~GHz by Morganti, Killeen, \& Tadhunter 1993) with
a bending angle of $\sim75^\circ$.
\\

\noindent   
{\bf 1519$-$273 (Fig.~2L )}
 
1519$-$273 is probably a BL Lac object because of its featureless spectrum
(V\'eron-Cetty \& V\'{e}ron 1993).
Its redshift is unknown, but has been assigned a lower limit of 0.2 by 
Stickel, Fried, \& K\"{u}hr (1993). 
It is highly polarized at optical wavelengths (Impey \& Tapia 1988, 1990).
It is also one of the GHz-peaked-spectrum radio sources 
(Gopal-Krishna, Patnaik, \& Steppe 1983).
 
The VLA observations at 1.4 and 5~GHz revealed no structure in 
1519$-$273 (Perley 1982).
Some sensitive intercontinental one-baseline VLBI surveys
(e.g. 2.3 and 8.4~GHz survey by
Morabito \etal 1986) suggested that 1519$-$273 has a very compact core.
The source was thought to be the most compact of all the 23 sources
detected by the TDRSS 2.3-GHz space VLBI observations (Linfield \etal 1989),
with a model size of 0.36~mas and an implied brightness temperature of 
3.0~$\times$~$10^{12}$~(1+z)~K.
 
Our observations yielded a first VLBI image of 1519$-$273.
From our map, it is unresolved with flux density of 2.2~Jy.
Its brightness temperature is greater than 3.2~$\times$$10^{12}$~K.
We propose that 1519$-$273 is unresolved and is a candidate 
calibrator for southern VLBI experiments at 5~GHz.
\\

\noindent 
{\bf 1548+056 (4C~05.64; No map)}
 
This source is a highly polarized quasar (Impey \& Tapia 1990) with a 
redshift of 1.442 (White \etal 1988), and is also classified as a 
blazar (Impey \& Tapia 1988).
In the radio band it shows low-frequency variability (Ghosh \& Rao 1992).
It is a core-dominated quasar with a flat spectrum at radio
wavelengths (Ghisellini \etal 1993).
We could not map 1548+056 due to the very limited data available.
The total flux density measured around 1992 was about 1.7~Jy, which is less
than the previous value of 3.3~Jy (e.g., Gregory \& Condon 1991).
The correlated flux density within Australia on baselines less than
3000~km is about 0.5~Jy, while it
was not detected on the baselines from Shanghai to Australia.
It probably has structure on the scale of about 4~mas.
\\

\noindent 
{\bf 1730$-$130 (NRAO~530; Fig.~2M )} 
 
This source is one of the best-known examples of a highly variable optical 
extragalactic source with a redshift of 0.902.
Its polarizations at both the optical and radio wavelengths, however,
are not large and thus 1730$-$130 is classified as a weakly polarized quasar.
As a low-frequency variable radio source, 1730$-$130 has been extensively 
monitored (e.g., at 327~MHz by Ghosh, Gopal-Krishna, \& Rao 1994; 
at 408~MHz by Bondi \etal 1994).
With VLBI observations at 10.7~GHz, Marscher \& Broderick
(1981) modeled 1730$-$130 as a double circular (0.31~mas in diameter) Gaussian
source separated about 1.2~mas along a position angle of $-156^\circ$, with the
flux densities of northeast and southwest components of
2.0~Jy and 1.6~Jy, respectively.
 
Romney \etal (1984) found that 
the structure of 1730$-$130 at 1.6~GHz is oriented in a 
north-south direction, 
extending about 26~mas at a P.A.~$-7^\circ$.
Based on two-epoch 1.6-GHz VLBI observations,
Padrielli \etal (1986) suggested that 1730$-$130 belongs to one of three
classes of low-frequency variable sources.
Its complex structural variations are not readily interpreted as angular expansion 
because the separation of components and the component sizes appear unchanged.
More likely, a change in the flux density
of a single component is responsible for the variation.

We fit our data with a three-component model (see Table 4).
It is clear from the map that there is a curvature of the components
relative to the central core, from $25^\circ$ at 1.3~mas (component 2) to
$19^\circ$ at 4.0~mas (component 3).
Furthermore, if we accept that these features are connected to the 
extended northerly emission on the 1.6-GHz maps of Padrielli {\it et al.}, which has
a position angle of $-7^\circ$ at a distance of about 26~mas to the core,
then the curvature continues as the jet moves outward. 
\\
 
\noindent
{\bf 1741$-$038 (OT$-$068; Fig.~2N )}
 
This source is a highly polarized quasar (Impey \& Tapia 1990)
with a redshift of 1.054 (White \etal 1988).
Some monitoring programs have revealed the presence of flux-density
fluctuations at both lower and higher frequencies.
The variability at 1.4 and 2.3~GHz was attributed
to RISS (Refractive Interstellar Scintillation)
(Hjellming \& Narayan 1986; Fiedler \etal 1987), 
while the variability at high frequency (15 and 22~GHz) was attributed
to the intrinsic change in the source (Hjellming \& Narayan 1986).
However, the flux density varies much less at 5~GHz. 
Quirrenbach \etal (1992) found intraday variability in 1741$-$038
with the Bonn 100-m telescope at 5~GHz.
They suggested that this scale of variability is preferentially 
found in those very compact VLBI sources.
 
Our VLBI image of 1741$-$038 shows a core-jet structure.
The results from model-fitting are listed in Table 4.
These are in close agreement with those reported from the study of
ESEs (Extreme Scattering Events) in 1741$-$038 (Fey, Clegg, \& Fiedler 1995).
The brightness temperature of the compact core is about $1.1\times10^{12}$~K,
consistent with the measurement of $0.9\times10^{12}$~K from the 
TDRSS VLBI experiment at 2.3~GHz (Linfield \etal 1989).
\\
 
\noindent 
{\bf 1814$-$637 (No map)}
 
This source is the southernmost one in our sample, at $\delta$ $\sim-64^\circ$,
beyond the southern limit of the Shanghai antenna.
It was identified as a Seyfert 2 galaxy with a redshift of 0.064
(Danziger \& Goss 1979; Thompson, Djorgovski, \& De Carvalho 1990).
It has a compact steep spectrum with a spectral 
index of $-0.75$ (K\"{u}hr \etal 1981).  
We failed to detect fringes on the baselines from Australia to 
South Africa.
Within Australia, 1814$-$637 had a correlated flux density of $\sim$1.1~Jy
on the shortest baseline of $\sim$100~km, and approximately 0.4~Jy on the baselines
of $\sim$1300~km.
This probably implies a flux density of only $\sim$1.0~Jy within a central 
$\sim$10~mas component.
No measurement of the total flux density was available
during the observations.
\\

\noindent
{\bf 1921$-$293 (OV$-$236; Fig.~2O )}
 
This object is the strongest source in our sample and has the lowest
redshift (z=0.352) in the sample of the sources imaged.
It is sometimes referred to as a highly polarized quasar
(Worrall \& Wilkes 1990), sometimes as an optically
violent variable quasar (Pica \etal 1988), and sometimes as a BL Lac
object (Litchfield, Robson, \& Stevens 1994).
It may be a blazar (Angel \& Stockman 1980),
since it exhibits variability at all observed wavelengths.
 
Because of the southern declination of 1921$-$293, 
few mapping observations have been carried out.
It was not resolved by the VLA at 5 and 1.4~GHz 
(de Pater, Schloerb, \& Johnson 1985; Perley 1982).
VLBI experiments at 2.3~GHz (Preston \etal 1989)
found that it was slightly resolved on baselines within Australia.
The TDRSS VLBI experiment at 2.3~GHz
(Linfield \etal 1989) estimated 
brightness temperature of $\sim$~$3.8\times10^{12}$~K for its compact core.
 
Our VLBI map is the first one at 5~GHz.
It clearly shows that 1921$-$293 has a strong compact core (component 1),
which contains flux density of 11.1~Jy at 5~GHz, or 77\% of the total flux 
density of about 14.4~Jy, within an area less than one quarter of the beam.
The parameters for the other two features are listed in Table 4. 
This morphology is possibly related to the core activity and jet curvature 
observed at 7~mm wavelength (Kellermann, private communication).
The quality of our data precluded an accurate estimation of the position
of component 2.
A full-track 5-GHz SHEVE observation in February 1993 showed a jet component 
along $30^\circ$, which is approximately $50^\circ$ different from what we
obtained for component 2.
This difference might indicate a precession of the axis of the jet,
and is certainly worthy of further investigation.
\\
 
\noindent
{\bf 2134+004 (OX+057; DA~553; PHL~61; Fig.~2P )}
 
This source was noted as an optical variable when it was identified
with a 17th-magnitude stellar object at a redshift of 1.94 
(Shimmins \etal 1968).
It is one of the most luminous objects in the universe.
Moore and Stockman (1984) classified it as a weakly polarized quasar.
Its radio spectrum peaks at about 5~GHz
(Stanghellini \etal 1990).
 
As a core-dominated radio source,
2134+004 is unresolved with the VLA at 5~GHz (Perley 1982).
On the parsec scale, it has a complex structure.
The earlier VLBI observations of 2134+004 were fit
with a collinear triple model at 10.7~GHz 
(Schilizzi \etal 1975) or a
two-component model at 5~GHz (Pauliny-Toth \etal 1981).
From 1987, however, Pauliny-Toth \etal (1990) noticed that the 
fine structure of 2134+004 has shown an increase in its extent, 
complexity and orientation, without any accompanying
radio outburst.
It is not clear how the new features are related to those identified
in 1970.

Our measurements at 5 GHz show that 2134+004 consists of three resolved
components: a strong compact core (component 1) with flux density of 6.7~Jy,
and two others (components 2 and 3) separated from component 1 by 2.5~mas at
P.A.~$-122^\circ$ and by 4.0~mas at P.A.~$-152^\circ$, respectively.
Components 2 and 3 both have flux densities of about 1.0~Jy.
Component 1 on our map may correspond to the strongest component in the
10.7-GHz map from 1989.27 by Pauliny-Toth \etal (1990).
The other two components in the 5-GHz map are probably related 
to the other part of the complex structure at 10.7~GHz.
The components appear to become opaque at different frequencies,
which makes the comparison of maps obtained at different wavelengths 
and different epochs difficult.
Multi-frequency VLBI monitoring would be extremely useful to
identify its core and understand the morphology.
It is possible that 
quasi-stationary shocks similar to those in 3C~454.3 are at work.
\\

\noindent 
{\bf 2145+067 (4C+06.69; OX+076.1; DA~562; Fig.~2Q )}
 
This is a weakly polarized quasar (Wills \etal 1992) at a redshift of 0.990.
At radio wavelengths it shows low-frequency variability (Mitchell \etal 1994).
Its spectral distribution  peaks at 15~GHz with flux density of
5.4~Jy (Gear \etal 1994).
 
The VLBI observations indicated that 2145+067 is very compact.
Linfield \etal (1990) detected it on the ground-space baseline
(1.38 earth diameters) at 15~GHz and modeled it with a circular Gaussian
component of diameter 0.16~mas and flux density of 5.43~Jy. 
This corresponds to a brightness temperature of $3.3\times10^{12}$~K.
With VLBI observations at 5~GHz in 1988.2,
Wehrle \etal (1992) found a strong, well-resolved component
elongated to the southeast,
and a diffuse component located at 7.1~mas with P.A.~$140^\circ$.
 
The sparse data for 2145+067 restricts the quality of our hybrid mapping.
No unique model could be found that fitted the data satisfactorily.
A simple model with only two components that fits our data has
an unresolved (about 0.2~mas in diameter) strong (4.0~Jy)
circular Gaussian component (1), and a secondary (1.2~Jy in flux density)
slightly resolved component (2).
The 4.0-Jy component probably corresponds to the component
detected from space
VLBI experiments, considering that 2145+067 is peaked at 15~GHz.
Thus its spectral index between 5~GHz and 15~GHz would be 0.3.
We assume that the stronger compact component 1 is the core.
The secondary compact component (2) is probably related to the 
diffuse component in the Wehrle \etal (1992) maps.
If these assignments are correct, a
jet curvature must occur, bending $\sim34^\circ$ 
from 1.2~mas to 7.1~mas.
\\

\noindent 
{\bf 2216$-$038 (4C$-$03.79; OY$-$027; Fig.~2R )}
 
A weakly polarized quasar (Impey \& Tapia 1990) with a redshift of 0.901,
2216$-$038 has a complex (or curved) radio spectrum (K\"{u}hr \etal 1981).
It sometimes shows low-frequency variation (Ghosh \& Rao 1992).
The VLA observations (Perley 1982) found a jet-like
feature extending to 8~arcseconds in P.A.~$140^\circ$, 
in addition to an unresolved core.
 
Our visibility data were well-modeled by a single 
circular Gaussian component, which has a flux density of 2.6~Jy 
and an angular size of 0.5~mas. 
Compared to the beam size (6.2~$\times$~0.9~mas at P.A.~$3^\circ$),
the source is not resolved. This is consistent with the TDRSS 2.3-GHz space
VLBI observations of Linfield \etal (1989), who estimated its size to be
0.5~mas with a flux density of 1.8~Jy at 2.3~GHz.
Therefore, the compact core has a spectral index of about 0.5 between
2.3 and 5.0~GHz.
\\

\noindent 
{\bf 2223$-$052 (3C~446; 4C$-$05.92; OY$-$039; NRAO~0687; Fig.~2S )}
 
This source is one of the most luminous quasars known and also one of
the most rapidly variable.
It has the properties of 
low-frequency variables (Padrielli \etal 1987), 
optically violent variables (Barbieri \etal 1990), and
highly polarized quasars (Moore \& Stockman 1981, 1984).
It is sometimes referred to as a BL Lac object (e.g.,
Miller \& French 1978).
The redshift of the source is 1.404 (Burbidge, Crowne, \& Smith 1977).
 
Our map of 2223$-$052 exhibits a distorted core-jet morphology,
which is in good agreement with many previous radio observations.
The source has a compact core and possibly two components to the east of the core.
Their parameters are given in Table 4.
Simon, Johnston, \& Spencer (1985) mapped the source at 1.6~GHz with VLBI
and found a continuous, slightly curved 
jet to the east, with a length of $\sim$ 250 mas.
The jet components seen on our map could be the knots consisting 
of relatively bright emission in this continuous jet.
The source was quiescent during our observations, with 
a measured total flux density of 5.4~Jy at 5~GHz.
Our model could account for no more than 75\% of single-dish measurement,
which implies that a quarter of the emission was resolved.
 
Based on all the observations available,
we suggest the following conclusion about the complex structure of 2223$-$052.
There is a very compact, optically thick core 
dominating the radio emission at higher frequencies.
At 100~GHz, it is 
less than 30~$\mu$as in diameter (Lerner \etal 1993).
This core is responsible for the extreme activity observed, which 
makes the determination of its spectral index difficult.
Second, there is an asymmetrical jet within the 2-arcsecond region, 
with some bright knots in it.
Beyond this jet there is possible diffuse emission, which was weaker than 
the 1.3~mJy per 4.4~arcseconds beam detected by Antonucci (1986).
The jet is initially ejected from the core at a P.A.~$-140^\circ$ at 
100$-$200~$\mu$as scale (Lerner \etal 1993), bends first 
clockwise to the north
to P.A.~$\sim90^\circ-100^\circ$ at 1$-$3~mas,
and then counterclockwise to the south to
P.A.~$110^\circ$ at 4$-$5~mas 
(Brown \etal 1981; Wehrle, Cohen, \& Unwin 1990a; this work).
The jet probably continues in a smoothly sinuous curved path
toward the east to a distance of $\sim550-650$~mas from the core
(Miley, Rickett, \& Gent 1967; Brown \etal 1981; 
Ulvestad, Johnston, \& Weiler 1983;
Simon, Johnston, \& Spencer 1985; Fejes, Porcas, \& Akujor 1992), 
where a sharp bending of $\sim120^\circ$ is required in order to align the 
jet with low-frequency arcsecond emission at P.A.~$-30^\circ$ on the
1$-$2 arcsecond-scale (Joshi \& Gopal-Krishna 1977; Browne \etal 1982).
Such a complex three dimensional structure reminds us that it could be
related to the helix in the jet if the axis is very close to the line of
sight.
\\

\noindent 
{\bf 2345$-$167 (OZ$-$176; Fig.~2T )}
 
This source is a highly polarized quasar (Moore \& Stockman 1981) with
a redshift of 0.576 (Hewitt \& Burbidge 1987).
It is an optically violent variable (Webb \etal 1988).
At radio frequencies, it has a complex spectrum with a peak around 5~GHz
(K\"{u}hr \etal 1981), and 
shows low-frequency variation (McAdam \& White 1983).
 
Our map of 2345$-$167 shows a core-jet structure.
The compact core (component 1) is barely resolved with 
a flux density of 1.6~Jy in an elliptical Gaussian component of 1.0~$\times$~0.7~mas at a
P.A. of $88^\circ$, which is in reasonable agreement with the model-fitting
results from the TDRSS 2.3-GHz space VLBI (Linfield \etal 1989).
Considering the estimated flux density of 1.3~Jy at 2.3~GHz from 
Linfield \etal (1989), 
we derive a spectral index of 0.28 for the core between 2.3 and 5~GHz.
The jet (component 2) is also resolved and separated by 3.0~mas 
from the core at 
a P.A. of $116^\circ$ with the flux density of 0.3~Jy.
Comparing our map with that from Wehrle \etal (1992) at 5~GHz,
we note that the strong secondary component in Wehrle \etal 
(1.5~mas from the core at P.A.~$\sim110^\circ$)
could be the same component as our jet.
Thus, we can estimate the proper motion of 0.26~mas~yr$^{-1}$,  
or an apparent velocity of $\sim$5.0c.
 
We suggest the following scenario for the jet's motion.
After being ejected from the core, the jet moves out along the direction of
$\sim110^\circ$ (Wehrle \etal 1992; this work) up to a distance of 5 mas, then turns
north to reach a P.A. of $25^\circ$ at 6~mas (Preston \etal 1989), 
and continues to bend to a P.A. of $\sim-5^\circ$ at 1.8 arcseconds
(Perley 1982).
It may then bend further clockwise on the scale of 4.0~arcseconds to a
P.A. of $-130^\circ$ (Wardle, Moore, \& Angel 1984).
\\

\section{Comparison with EGRET Detection }

Fifty-one active galactic nuclei have been detected above 100~MeV
by EGRET on board CGRO (Thompson \etal 1995).
These $\gamma$-ray-loud sources are preferentially flat-spectrum,
compact, radio-loud sources.
Thus, the study of their radio properties is of great importance for our
understanding the process of $\gamma$-ray emission.

Of our twenty-two sources, five have $\gamma$-ray emission detected 
by EGRET (von Montigny \etal 1995).
These are 0235+164, 0420$-$014, 0458$-$020, 1510$-$089 and 1741$-$038.
Of these, 1741$-$038 does not appear in the second EGRET catalog (Thompson
\etal 1995) because it was detected with statistical significance (3.9)
just below the threshold (4.0) for the new catalog.
We retain it as a $\gamma$-ray source in the discussion.
Superluminal motion was detected in one source ($\beta_{\small app}$=3.9c 
in 0420$-$014), as discussed in $\S4$.
One (0235+164) is a BL Lac object, and the remaining four are highly polarized
quasars (HPQs).
In other words, HPQs are much more easily detected at $\gamma$-ray
wavelength than any other types.
Since HPQs are thought to beam their radio emissions towards
observers, we could infer from the close relation between HPQs and $\gamma$-ray
sources that a kind of similar $\gamma$-ray beaming effect is at work,
although both emissions could not originate in the same region.

With VLBI maps available for twenty sources, we can obtain their structural
parameters and calculate the brightness temperature (T$_{\small B}$), as
shown in Table 4.
In Figure 3, we plot the distribution of T$_{\small B}$
(or its lower limit) for both the 
five $\gamma$-ray sources and the other fifteen sources.
Fig.~3(a) is a histogram of the brightest components in the twenty sources
(i.e., those labeled component 1 in Table 4), while Fig.~3(b) includes all
forty-five model components in Table 4.
The black areas correspond to the components in the sources detected by
EGRET, while the unfilled areas represent the others.
For those unresolved components, the lower limits of T$_{\small B}$ are
applied. From Fig.~3(a), the fraction of sources having T$_{\small B}$ exceeding 
10$^{12}$~K in sources with $\gamma$-ray emission detected is 4/5, which
is the same as that in the remaining sources (12/15).
If we exclude 1741$-$038 from the $\gamma$-ray source list, such a ratio would
be 3/4 for sources with $\gamma$-ray emission and 13/16 for the others.
Comparing Fig.~3(b) with Fig.~3(a), we can see a large range of T$_B$ for the
jet-like components, while the distribution of T$_B$ for the strong compact core
is around the inverse Compton Limit ($\sim$10$^{12}$K) with a small spread. The
jet components might have less connection to the $\gamma$-ray emission.
The relativistic beaming effect can be invoked to explain brightness
temperatures in excess of the Compton Limit. Therefore, the similarity of T$_B$
distributions in Fig.~3(a), i.e., the fact that more than 75\% have T$_B$ greater
than 10$^{12}$K in the subsamples of both $\gamma$-ray and
non-$\gamma$-ray sources, might
be an indication of the existence of beamed $\gamma$-ray emission too.
Furthermore, considering the fact that most $\gamma$-ray sources are strongly
variable, the similarity and strong correlation of $\gamma$-ray sources
with radio-loud sources such as HPQs may indicate that all the strong
radio variables will be detected in $\gamma$-ray emission if they are
observed at a suitable time and with small angles to the ejection.
\\

\section{Summary }
 
In this paper we have presented southern 5-GHz VLBI observations of
compact radio structure in a sample of twenty-two extragalactic radio sources.
The source sample with selection criteria, 
observations and data reduction were described.
Most of these sources show strong variability at the radio wavelengths.
This is the first VLBI imaging survey carried out at 5~GHz for
equatorial and southern radio sources.
Our work redresses the absence of information at frequencies 
between 2.3~GHz and 8.3~GHz.
 
Generally, the quality of the maps depends on the (u,v) coverage,
sensitivity, and uncertainties in the calibration.
Our examples show that we could obtain the basic information for most sources 
even with snapshot-mode observations.
Although we were forced by the limited data to favor the simplest models,
we believe that most sources in our sample were well represented by them.
 
The main conclusions drawn in this paper are:
 
(1) 
We have detected fringes from all twenty-two sources on almost all the baselines.
We mapped twenty of the twenty-two sources detected.
Of these twenty sources, six (0048$-$097, 0104$-$408, 1144$-$379, 1519$-$273,
1741$-$038 and 2216$-$038)
have never been mapped before.
Fifteen of the sources show a core-jet structure, 
one (1510$-$089) may have a two-sided jet and 
four (0104$-$408, 1144$-$379, 1519$-$273 and 2216$-$038) have a single compact
component.
There are no compact doubles, which is consistent with the fact 
that few compact
double sources show variability.
The new information for these southern sources will
be helpful in determining targets for future space VLBI missions. 
 
(2) 
The curvature of the jet
(i.e. the significant change in the position angle of jet components)
from mas- to arcsecond-scale, sometimes between
a few mas to 10 or 100~mas scale, seems to be a common feature.
We found eleven examples (0106+013, 0458$-$020, 1104$-$445, 1334$-$127,
1510$-$089, 1730$-$130, 1921$-$293, 2134+004, 2145+067,
2223$-$052 and 2345$-$167) in the total of sixteen core-jet sources.
 
(3)
We compared our maps to previous ones and estimated the ejection time from
peaks in the flux-density curves at radio wavelengths.
We confirmed the superluminal speed of 8.2c for one component, 
and suggested a new superluminal motion of 7.4c for another component
in 0106+013.
The superluminal motions were inferred for three other sources 
(3.9c in 0420$-$014, 5.2c in 1334$-$127 and 5.0c in 2345$-$167).
These values are typical of superluminal speeds in the strong, variable
quasars.
The results need to be confirmed by future VLBI observations.
 
(4)
Further VLBI observations of southern radio sources would be 
important not only for expanding the source sample, but for confirming 
the observed structures (such as bending and superluminal motion) and providing
multi-epoch observations, which are necessary for the study of variability. 

(5)
There are five sources (0235+164, 0420$-$014, 0458$-$020, 1510$-$089
and 1741$-$038) having $\gamma$-ray emissions detected by 
EGRET (von Montigny \etal 1995).
Analysis reveals the dominance of highly polarized quasars among the
$\gamma$-ray sources.
Comparison of VLBI measurements with EGRET detection shows  
a similarity of T$_{\small B}$ distribution between these
$\gamma$-ray sources and others.
Taken together, these imply the existence of the beamed $\gamma$-ray emission.

\acknowledgments
The authors wish to thank H. D. Aller and M. F. Aller for data taken by
the University of Michigan Radio Astronomy Observatory (UMRAO) at 4.8, 8.0 and 
14.5~GHz; 
E.~\null Valtaoja and H.~Ter\"{a}sranta for data taken at the Mets\"{a}hovi
Radio Research Station at 22 and 37~GHz; 
and E.~Valtaoja and M.~Tornikoski
for data taken at the Swedish-ESO Submillimeter Telescope (SEST) at 90
and 230~GHz, prior to publication.

This work was supported at Shanghai Astronomical Observatory by grants from
the National Program for the Enhancement of Fundamental Research.
Part of this research was carried out at the Jet Propulsion
Laboratory, California Institute of Technology, under
contract with the National Aeronautics and Space Administration.
Z.-Q. Shen benifited from his discussions with M. Reid, M. Birkinshaw
and C. Carilli.
Z.-Q. Shen acknowledges the receipt of a Smithsonian Pre-doctoral Fellowship.

\clearpage

\begin{center}
{\bf \large Figure Captions}
\end{center}

\figcaption{The typical (u,v) coverage for the SHEVE plus the 
	Shanghai station snapshot observations (1730$-$130). \label{fig1}}

\figcaption{The VLBI maps of twenty extragalactic radio sources 
	observed in November 1989. The synthesized beam is shown in the 
	lower left of each map. See Table 3 for detailed parameters. \label{fig2}}

\figcaption{The histograms of the brightness temperatures, 
	(a) for brightest component in twenty sources mapped
	and (b) for all forty-eight components modeled, as shown in Table 4. 
	The black areas correspond to the components in the five sources 
	detected at $\gamma$-ray wavelength,
	while the unfilled areas represent those in the other fifteen sources. \label{fig3}}

\clearpage


\clearpage

\begin{thebibliography}{}
\bibitem{}Andrew, J. P., \& Smith, A. G. 1983, \apj, 272, 11
\bibitem{}Angel, J. R. P., \& Stockman, H. S.  1980, \araa, 18, 321
\bibitem{}Antonucci, R. R. J., \& Ulvestad, J. S. 1985, \apj, 294, 158
\bibitem{}Antonucci, R. R. J. 1986, \apj, 304, 634
\bibitem{}B{\aa}{\aa}th, L. B. 1984, in IAU Symp. 110, VLBI and Compact Radio Sources, 
	edited by R. Fanti, K.~I.~Kellermann, and G. Setti (Reidel, Dordrecht), p. 127
\bibitem{}Barbieri, C., Vio, R., Cappellaro, E., \& Turatto,~M. 1990, \apj, 359, 63
\bibitem{}Barrs, J. W. M., Genzel, R., Pauliny-Toth, I. I. K., \& Witzel, A. 1977,
	\aap, 61, 99
\bibitem{}Barthel, P. D., \& Miley, G. 1988, \nat, 333, 319
\bibitem{}Bondi, M., Padrielli, L., Gregorini, L., Mantovani, F.,
	Shapirovskaya, N., \& Spangler, S. R. 1994, \aap, 287, 390
\bibitem{}Briggs, F. H., Wolfe, A. M., Liszt, H. S., Davis, M. M., \&
	Turner, K. L. 1989, \apj, 341, 650
\bibitem{}Brown, R. L., Johnston, K. J., Briggs, F. H., Wolfe, A. M., 
	Neff, S. G., \& Walker, R. C. 1981, \aplett, 21, 105
\bibitem{}Browne, I. W. A., {\it et al.}, 1982, \nat, 299, 788
\bibitem{}Burbidge, E. M. 1966, \apj, 143, 612
\bibitem{}Burbidge, E. M., \& Kinman, T. D. 1966, \apj, 145, 654
\bibitem{}Burbidge, E. M., Caldwell, R. D., Smith, H. E., Liebert, J., \&
	Spinrad, H. 1976, \apjl, 205, L117
\bibitem{}Burbidge,~G.~R., Crowne,~A.~H., \& Smith,~H.~E. 1977, \apjs, 33, 113
\bibitem{}Cohen,~R.~D., Smith,~H.~E., \& Burbidge,~E.~M. 1986, \baas, 18, 674
\bibitem{}Costa,~E., \& Loyola,~P. 1992, \aaps, 96, 183
\bibitem{}Danziger,~I.~J., \& Goss,~W.~M. 1979, \mnras, 186, 93
\bibitem{}de Pater,~I., Schloerb,~F.~P., \& Johnson,~A.~H. 1985, \aj, 90, 846
\bibitem{}de Waard,~G.~J. 1986, in
	Thermal-Nonthermal Relationships in Active Galactic Nuclei
	(Sterrewacht, Leiden), p. 73
\bibitem{}Falomo,~R., Melnick,~J., \& Tanzi,~E.~G. 1990, \nat, 345, 692
\bibitem{}Fejes,~I., Porcas,~R.~W., \& Akujor,~C.~E. 1992, \aap, 257, 459
\bibitem{}Fey,~A., Clegg,~A., \& Fiedler,~R. 1995, BAAS, 27, 863
\bibitem{}Fiedler,~R.~L., {\it et al.}, 1987, \apjs, 65, 319
\bibitem{}Gabuzda,~D.~C., Cawthorne,~T.~V., Roberts,~D.~H., \&  Wardle,~J.~F.~C.
	1992, \apj, 388, 40
\bibitem{}Gear,~W.~K., {\it et al.}, 1994, \mnras, 267, 167
\bibitem{}Ghisellini,~G., Padovani,~P., Celotti,~A., \& Maraschi,~L. 1993, \apj, 407, 65
\bibitem{}Ghosh,~T., \& Rao,~A.~P. 1992, \aap, 264, 203
\bibitem{}Ghosh,~T., Gopal-Krishna, \& Rao,~A.~P. 1994, \aaps, 106, 29
\bibitem{}Gopal-Krishna, Patnaik,~A.~R., \& Steppe,~H. 1983, \aap, 123, 107
\bibitem{}Gregory,~P.~C., \& Condon,~J.~J. 1991, \apjs, 75, 1011
\bibitem{}Henstock,~D.~R., {\it et al.}, 1995, \apjs, 100, 1
\bibitem{}Hewitt,~A., \& Burbidge,~G. 1987, \apjs, 63, 1
\bibitem{}Hirosawa, H. 1991, in Frontiers of VLBI, edited by H.~Hirabayashi,
          M.~Inoue, and H.~Kobayashi (Universal Academy Press, Tokyo), p. 21
\bibitem{}Hjellming,~R.~M., \& Narayan,~R. 1986, \apj, 310, 768
\bibitem{}Impey,~C.~D., \& Tapia,~S. 1988, \apj, 333, 666
\bibitem{}Impey,~C.~D., \& Tapia,~S. 1990, \apj, 354, 124
\bibitem{}Jauncey,~D.~L., {\it et al.}, 1989, \aj, 98, 44
\bibitem{}Jauncey,~D.~L., {\it et al.}, 1994, in IAU Symp.~158, Very High Resolution
        Imaging, edited by J.~G.~Roberston and W.~J.~Tango
        (Kluwer, Dordrecht), p. 131
\bibitem{}Jones,~D.~L., B{\aa}{\aa}th,~L.~B., Davis,~M.~M., \& Unwin,~S.~C.
	1984, \apj, 284, 60
\bibitem{}Joshi,~M.~N., \& Gopal-Krishna 1977, \mnras, 178, 717
\bibitem{}Kardashev,~N.~S., \& Slysh,~V.~I. 1988, in IAU Symp.~129, The Impact 
	of VLBI on Astrophysics and Geophysics, edited by M.~J. Reid and J.~M. Moran
        (Kluwer, Dordrecht), p. 433
\bibitem{}Kollgaard,~R.~I., Wardle,~J.~F.~C., \& Roberts,~D.~H. 1990, \aj, 100, 
	1057
\bibitem{}K\"{u}hr,~H., Witzel,~A., Pauliny-Toth,~I.~I.~K., \& Nauber,~U.
	1981, \aaps, 45, 367
\bibitem{}Lerner,~M.~S., {\it et al.}, 1993, \aap, 280, 117
\bibitem{}Linfield,~R.~P., {\it et al.}, 1989, \apj, 336, 1105
\bibitem{}Linfield,~R.~P., {\it et al.}, 1990, \apj, 358, 350
\bibitem{}Litchfield,~S.~J., Robson,~E.~J., \& Stevens,~J.~A. 1994, \mnras, 270, 341
\bibitem{}Marscher,~A.~P., \& Broderick,~J.~J. 1981, \apj, 249, 406
\bibitem{}McAdam,~W.~B. 1982, in
	Proc. Workshop on Low Frequency Variability of Extragalactic
	Radio Sources, edited by W.~D. Cotton \& S.~R. Spangler
	(NRAO, Green Bank), p. 141
\bibitem{}Meier,~D.~L., {\it et al.}, 1989, \aj, 98, 27
\bibitem{}Miley,~G.~K., Rickett,~B.~J., \& Gent,~H. 1967, \nat, 216, 974
\bibitem{}Miller,~J.~S., \& French,~H.~B. 1978, in
	Pittsburgh Conference On BL Lac Objects, Department of
        Physics and Astronomy, University of Pittsburgh, edited by A.~M. Wolfe, 228
\bibitem{}Mitchell,~K.~J., {\it et al.}, 1994, \apjs, 93, 441
\bibitem{}Moore,~R.~L., \& Stockman,~H.~S. 1981, \apj, 243, 60
\bibitem{}Moore,~R.~L., \& Stockman,~H.~S. 1984, \apj, 279, 465
\bibitem{}Morabito,~D.~D., Niell,~A.~E., Preston,~R.~A., Linfield,~R.~P.,
	Wehrle,~A.~E., \& Faulkner,~J. 1986, \aj, 91, 1038
\bibitem{}Morganti,~R., Killeen,~N.~E.~B., \& Tadhunter,~C.~N. 1993, \mnras, 263, 1023
\bibitem{}Neff,~S.~G., \& Hutchings,~J.~B. 1990, \aj, 100, 1441
\bibitem{}Nicolson,~G.~D., Glass,~I.~S., Feast,~M.~W., \& Andrews,~P.~J.
	1979, \mnras, 189, 29p
\bibitem{}O'Dea,~C.~P., Barvainis,~R., \& Challis,~P.~M. 1988, \aj, 96, 435
\bibitem{}Padrielli,~L., {\it et al.}, 1986, \aap, 165, 53
\bibitem{}Padrielli,~L., {\it et al.}, 1987, \aaps, 67, 63
\bibitem{}Pauliny-Toth,~I.~I.~K., Preuss,~E., Witzel,~A., Graham,~D.,
	Kellermann,~K.~I., \& R\"{o}nn\"ang,~B. 1981, \aj, 86, 371
\bibitem{}Pauliny-Toth,~I.~I.~K., Zensus,~A., Cohen,~M.~H., Alberdi,~A.,
	\& Shaal,~R. 1990, in Parsec-Scale Radio Jets, 
	edited by J.~A.~Zensus and T.~J.~Pearson (Cambridge University
	Press, Cambridge), p. 55
\bibitem{}Pearson,~T.~J., \& Readhead,~A.~C.~S. 1981, \apj, 248, 61
\bibitem{}Pearson,~T.~J., \& Readhead,~A.~C.~S. 1988, \apj, 328, 114
\bibitem{}Perley,~R.~A. 1982, \aj, 87, 859  
\bibitem{}Pica, A. J., Smith, A. G., Webb, J. R., Leacock, R. J., Clements, S.,
        \& Gombola, P. P. 1988, \aj, 96, 1215
\bibitem{}Polatidis,~A.~G., {\it et al.}, 1995, \apjs, 98, 1
\bibitem{}Pollock,~J.~T., Pica,~A.~J., Smith,~A.~G., Leacock,~R.~J., 
	Edwards,~P.~L., \& Scott,~R.~L. 1979, \aj, 84, 1658
\bibitem{}Preston,~R.~A., Morabito,~D.~D., Williams,~J.~G., Faulkner,~J., 
	Jauncey,~D.~L., \& Nicolson,~G.~D 1985, \aj, 90, 1599
\bibitem{}Preston,~R.~A., {\it et al.}, 1989, \aj, 98, 1
\bibitem{}Price,~R., Gower,~A.~C., Hutchings,~J.~B., Talon,~S., Duncan,~D., \&	Ross,~G.
	1993, \apjs, 86, 365
\bibitem{}Quiniento,~Z.~M., Cersosimo,~J.~C., \& Colomb,~F.~R. 1988, \aaps, 76, 21
\bibitem{}Quirrenbach,~A., {\it et al.}, 1992, \aap, 258, 279
\bibitem{}Radecke,~H.-D., {\it et al.}, 1995, \apj, 438, 659
\bibitem{}Rieke,~G.~H., Grasdalen,~G.~L., Kinman,~T.~D., Hintzen,~P.,
	Wills,~B.~J., \& Wills,~D. 1976, \nat, 260, 754
\bibitem{}Romney,~J., {\it et al.}, 1984, \aap, 135, 289
\bibitem{}Schilizzi,~R.~T., {\it et al.}, 1975, \apj, 201, 263
\bibitem{}Shepherd,~M.~C., Pearson,~T.~J., \& Taylor,~G.~B. 1994, \baas, 26, 987
\bibitem{}Shimmins,~A.~J., Searle,~L., Andrew,~B.~H., \& Brandie,~G.~W. 1968,
	\aplett, 1, 167
\bibitem{}Simon,~R.~S., Johnston,~K.~J., \& Spencer,~J.~H. 1985, \apj, 290, 66
\bibitem{}Spinrad,~H., \& Smith,~H.~E. 1975, \apj, 201, 275
\bibitem{}Stanghellini,~C., O'Dea,~C.~P., Baum,~S.~A., \& Fanti,~R. 1990, in
        Compact Steep-Spectrum and GHz-Peaked Spectrum Radio Sources, 
	edited by C.~Fanti, R.~Fanti, C.~P. O'Dea, and R.~T. Schilizzi
	(Instituto di Radioastronomia, Bologna), p. 55
\bibitem{}Steppe,~H., ~H., Salter,~C.~J., Chini,~R., Kreysa,~E., Brunswig,~W., \& 	
	P\'{e}rez,~J.~L. 1988, \aaps, 75, 317
\bibitem{}Steppe,~H., Liechti,~S., Mauersberger,~R., K\"{o}mpe,~C.,
        Brunswig,~W., \& Ruiz-Moreno,~M. 1992, \aaps, 96, 441
\bibitem{}Stickel,~M., Fried,~J.~W., \& K\"{u}hr,~H. 1989, \aaps, 80, 103
\bibitem{}Stickel,~M., Padovani,~P., Urry,~C.~M., Fried,~J.~W., \& K\"{u}hr,~H.
	1991, \apj, 374, 431
\bibitem{}Stickel,~M., Fried,~J.~W., \& K\"{u}hr,~H. 1993, \aaps, 98, 393
\bibitem{}Takahashi,~Y., \& Kurihara,~N. 1993, \pasj, 45, 497 
\bibitem{}Taylor,~G.~B., {\it et al.}, 1994, \apjs, 95, 345
\bibitem{}Thakkar,~D.~D., {\it et al.}, 1995, \apjs, 98, 33
\bibitem{}Thompson,~D.~J., Djorgovski,~S., \& De Carvalho,~R. 1990, \pasp, 102, 1235
\bibitem{}Thompson,~D.~J., {\it et al.}, 1993, \apjl, 415, L13
\bibitem{}Thompson,~D.~J., {\it et al.}, 1995, \apjs, 101, 259
\bibitem{}Tornikoski,~M., Valtaoja,~E., Ter\"{a}sranta,~H., Lainela,~M.,
	Bramwell,~D., \& Botti,~L.~C.~L. 1993, \aj, 105, 1680
\bibitem{}Tzioumis,~A.~K., {\it et al.}, 1989, \aj, 98, 36
\bibitem{}Ulvestad,~J.~S., Johnston,~K.~J., \& Weiler,~K.~W. 1983, \apj, 266, 18
\bibitem{}Vermeulen,~R.~C. \& Cohen,~M.~H. 1994, \apj, 430, 467
\bibitem{}V\'eron-Cetty,~M.-P., \& V\'eron,~P. 1993, \aaps, 100, 521
\bibitem{}von Montigny,~C., {\it et al.}, 1995, \apj, 440, 525
\bibitem{}Wagner,~S.~J., {\it et al.}, 1995, \aap, 298, 688
\bibitem{}Wardle,~J.~F.~C., Moore,~R.~L., \& Angel,~J.~R.~P. 1984, \apj, 279, 93
\bibitem{}Webb,~J.~R., Smith,~A.~G., Leacock,~R.~J., Fitzgibbons,~G.~L.,
	Gombola,~P.~P., \& Shepherd,~D.~W. 1988, \aj, 95, 374
\bibitem{}Wehrle,~A.~E., Cohen,~M.~H., \& Unwin,~S.~C. 1990a, in
        Parsec-Scale Radio Jets, edited by J.~A.~Zensus and T.~J.~Pearson (Cambridge
	University Press, Cambridge), p. 49
\bibitem{}Wehrle,~A.~E., Cohen,~M.~H., \& Unwin,~S.~C. 1990b, \apjl, 351, L1
\bibitem{}Wehrle,~A.~E., Cohen,~M.~H., Unwin,~S.~C., Aller,~H.~D., Aller,~M.~F.,
	\& Nicolson,~G. 1992, \apj, 391, 589
\bibitem{}Weiler,~K.~W., \& Johnston,~K.~J. 1980, \mnras, 190, 269
\bibitem{}White,~G.~L., {\it et al.}, 1988, \apj, 327, 561
\bibitem{}Wilkes,~B.~J. 1986, \mnras, 218, 331
\bibitem{}Wills,~B.~J., Wills,~D., Breger,~M., Antonucci,~R.~R.~J., \&
	Barvainis,~R. 1992, \apj, 398, 454
\bibitem{}Worrall,~D.~M., \& Wilkes,~B.~J. 1990, \apj, 360, 396
\bibitem{}Xu,~W., Readhead,~A.~C.~S., Pearson,~T.~J., Polatidis,~A.~G., \&
	Wilkinson,~P.~N. 1995, \apjs, 99, 297
\end{thebibliography}
\end{document}